\documentclass{mrm2plain}
\usepackage{siunitx}

\graphicspath{{./figures/}}

\papertype{Note}
\title{Robust Retrospective Frequency and Phase Correction for Single-voxel MR Spectroscopy}

\author[1]{Martin Wilson}

\affil[1]{Centre for Human Brain Health and School of Psychology, University of Birmingham, Birmingham, UK.}

\corraddress{Martin Wilson, Centre for Human Brain Health, University of Birmingham, Edgbaston, Birmingham, B15 2TT, United Kingdom.}

\corremail{wilsonmp@bham.ac.uk}

\runningauthor{Wilson}

\newcommand{\proton}{\ensuremath{^1\mathrm{H}}}
\newcommand{\bzero}{\ensuremath{\mathrm{B}_0}}

\begin{document}

\maketitle

\begin{abstract}
\noindent \textbf{Purpose:} subject motion and static field (\bzero) drift are known to reduce the quality of single voxel MR spectroscopy data due to incoherent averaging. Retrospective correction has previously been shown to improve data quality by adjusting the phase and frequency offset of each average to match a reference spectrum. In this work, a new method (RATS) is developed to be tolerant to large frequency shifts (greater than 7Hz) and baseline instability resulting from inconsistent water suppression. \\
\textbf{Methods:} in contrast to previous approaches, the variable-projection method and baseline fitting is incorporated into the correction procedure to improve robustness to fluctuating baseline signals and optimization instability. RATS is compared to an alternative method, based on time-domain spectral registration (TDSR), using simulated data to model frequency, phase and baseline instability. In addition, a J-difference edited glutathione in-vivo dataset is processed using both approaches and compared. \\
\textbf{Results:} RATS offers improved accuracy and stability for large frequency shifts and unstable baselines. Reduced subtraction artifacts are demonstrated for glutathione edited MRS when using RATS, compared with uncorrected or TDSR corrected spectra. \\
\textbf{Conclusion:} the RATS algorithm has been shown to provide accurate retrospective correction of SVS MRS data in the presence of large frequency shifts and baseline instability. The method is rapid, generic and therefore readily incorporated into MRS processing pipelines to improve lineshape, SNR and aid quality assessment.
\keywords{RATS, VARPRO, MRS, SVS, motion, drift}
\end{abstract}

% MRM abstract max length is 250 words
% MRM note max length is 2800 words, 5 fig plus tables

\vspace{1.0cm}
{\centering \textbf{Submitted to Magnetic Resonance in Medicine} \par}
\newpage

\section{Introduction}
Single voxel acquisition is currently the most widely used in-vivo \proton\ Magnetic Resonance Spectroscopy (MRS) technique for clinical brain investigation \cite{Oz2014}. Repeated acquisitions, known as averages or shots, are usually combined to attain a sufficient signal-to-noise ratio (SNR) to measure key metabolite signals, where a doubling the number of averages theoretically improves the SNR by a factor of $\sqrt{2}$. A typical acquisition protocol of 128 averages, acquired from a 2cm sided cubic volume, may be used to discern the major \proton\ metabolite resonances, such as total-N-acetylaspartate (tNAA), total-creatine (tCr), total-choline (tCho) and myo-inositol.

Ideally, metabolite signals are completely stable throughout the entire acquisition period - achieving the highest possible SNR when averaged. However, two primary mechanisms have been shown to result in dynamic perturbations in spectral phase and frequency. Firstly, slowly varying changes in the static field strength (\bzero\ drift) commonly follow gradient intensive sequences, such as echo-planer imaging, due to a heating of the static shim elements \cite{Foerster2005}. \bzero\ drift during MRS acquisition causes a slowly varying frequency offset, where subsequent averages become increasingly misaligned relative to the first. The second primary cause of temporal instability originates from subject movement, typically resulting in a transient change in the frequency offset and spectral phase.

Both \bzero\ drift and subject movement degrade the SNR and lineshape of MRS data due to incoherent averaging. Unstable acquisitions are particularly detrimental to J-difference edited experiments \cite{Mescher1998}, since spectral misalignment results in an incomplete subtraction of non-edited resonances - resulting in a distortion of the edited metabolite signals \cite{Evans2013}. One of the earliest approaches for retrospective MRS instability correction is based on a frequency and phase measurement made from the residual water signal \cite{Zhu1992}. The change in these parameters is estimated throughout the acquisition period, and each average corrected to obtain consistent spectra. However, the use of a residual water resonance has the disadvantage of potentially biasing the metabolite estimates through baseline distortion from the water resonance peak ``tails'', or sideband artifacts that increase with the residual water amplitude \cite{Clayton2001}.

More recently, a method has been developed to align spectra without the requirement for a residual water signal \cite{Near2015}. Correction is performed using a least-squares optimization to a reference spectrum in the time-domain, a process known as ``spectral registration''. The time-domain spectral registration (TDSR) approach has been compared with two other correction methods, based on the residual water signal \cite{Helms2001} and a metabolite peak fitting method \cite{Waddell2007}, and found to perform favorably.

In this paper, a new method for spectral registration is presented. In contrast to previous approaches, the registration problem is formulated as variable-projection (VARPRO) \cite{Golub1973, VanderVeen1988} in the frequency domain. The use of VARPRO allows the incorporation of baseline modeling, whilst also reducing the iterative optimization complexity from two parameters (phase and frequency) to one (frequency). The approach is compared with TDSR, and found to be more robust to large frequency shifts ($>$7Hz), baseline distortions and edited-MRS frequency misalignment.

\section{Methods}
\subsection{Time-domain spectral registration}
The TDSR method applies a frequency and phase adjustment to each target average, $\mathbf{S}(t)$, to match a reference signal, $\mathbf{R}(t)$ using nonlinear least-squares optimization. The optimization problem may be stated as:

\begin{equation}
	\min_{F \in \mathbb{R}, \thinspace \phi \in \mathbb{R}} \sum^{t_{N-1}}_{t=t_0} \bigg\lvert \mathrm{Re}(\mathbf{R}(t) - \mathbf{G}(t,F,\phi)) + \mathrm{Im}(\mathbf{R}(t) - \mathbf{G}(t,F,\phi)) \bigg\rvert ^2,
\label{tdsr}
\end{equation}

where $F$ is the frequency correction parameter in Hz, $\phi$ is the phase correction parameter in degrees and $\mathbf{G}(t,F,\phi)$ is defined as:

\begin{equation}
    \mathbf{G}(t,F,\phi)=\mathbf{S}(t) \thinspace e^{2 \pi j \left (Ft+\frac{\phi}{360} \right )}.
\label{corr_eqn}
\end{equation}

Whilst the parametric correction (\ref{corr_eqn}) is performed as a complex operation ($j=\sqrt{-1}$), the optimization problem (\ref{tdsr}) is real valued, achieved by the concatenation of real (Re) and imaginary (Im) parts of $\mathbf{R}$ and $\mathbf{G}$. The optimum $F$ and $\phi$ parameters are found using a non-linear least-squares regression algorithm and applied to the target average, $\mathbf{S}(t)$, to generate a corrected spectrum.

\subsection{Frequency-domain spectral registration with variable-projection baseline modeling}
One potential limitation of the TDSR method is the assumption that each average may be accurately matched to the reference signal by adjusting only the frequency and phase. Acquisitions with moderate residual water commonly exhibit baseline artifacts, which have a smooth appearance in the frequency domain, and often change throughout the acquisition period due to scanner instability or subject movement. In this paper, a modification of the TDSR optimization problem (\ref{tdsr}) is presented, incorporating baseline differences between the target and reference spectrum:

\begin{equation}
\begin{aligned}
    \min_{\substack{F \in \mathbb{R}, \thinspace \mathbf{a}_\mathrm{G} \in \mathbb{C} \\ \mathbf{a}_\mathrm{B} \in \mathbb{C}^{P+1}}} \sum^{f_{N-1}}_{f=f_0} \bigg\lvert & \mathrm{Re}(\hat{\mathbf{R}}(f) - \hat{\mathbf{G}}(f,F) \thinspace \mathbf{a}_\mathrm{G} + \mathbf{B}(f) \thinspace \mathbf{a}_\mathrm{B} ) \\ & + \mathrm{Im}(\hat{\mathbf{R}}(f) - \hat{\mathbf{G}}(f,F) \thinspace \mathbf{a}_\mathrm{G} + \mathbf{B}(f) \thinspace \mathbf{a}_\mathrm{B}) \bigg\rvert ^2.
\end{aligned}
\end{equation}

In contrast to TDSR, the objective function is in the frequency domain, where $\hat{\mathbf{R}}$ and $\hat{\mathbf{G}}$ represent $\mathbf{R}$ and $\mathbf{G}$ following Fourier transformation. A second modification is the addition of the complex amplitude parameter applied to target spectrum, $\mathbf{a}_\mathrm{G}$. Finally, a polynomial basis $\mathbf{B}$, scaled by $\mathbf{a}_\mathrm{B}$, is added to account for baseline differences between the target and reference spectrum. $\mathbf{B}$ is structured as basis matrix with $N$ rows and $p+1$ columns, where $N$ is the number of points considered in the frequency domain and $p$ represents the highest order basis polynomial: $\{1,x,x^2,...,x^{p}\}$. Concatenating the adjusted target spectrum, polynomial basis and corresponding complex amplitude parameters leads to:

\begin{equation}
    \hat{\mathbf{G}}_\mathrm{B} = \begin{bmatrix} \hat{\mathbf{G}} & \mathbf{B} \end{bmatrix},
\end{equation}

\begin{equation}
    \mathbf{a} = 
    \begin{bmatrix} 
        \mathbf{a}_\mathrm{G} \\ \mathbf{a}_\mathrm{B}
    \end{bmatrix}, 
\end{equation}

\begin{equation}
    \min_{F \in \mathbb{R}, \thinspace \mathbf{a} \in \mathbb{C}^{P+2}} \lVert \hat{\mathbf{R}} - \hat{\mathbf{G}}_\mathrm{B}(F) \thinspace \mathbf{a} \rVert_2^2,
\label{varpro}
\end{equation}

where the linear parameters $\mathbf{a}$ are separated from the non-linear frequency adjustment parameter $F$. The purpose of this reformulation is to  allow the solution of (\ref{varpro}) using the VARPRO approach, which has been shown to be particularly effective for solving similar problems, \cite{Golub1973, VanderVeen1988}. VARPRO exploits the fact that the linearly appearing parameters, $\mathbf{a}$, may be optimally found using stable and efficient linear methods:

\begin{equation}
    \min_{F \in \mathbb{R}} \lVert \hat{\mathbf{R}} - \hat{\mathbf{G}}_\mathrm{B}(F) \thinspace \hat{\mathbf{G}}_\mathrm{B}(F)^{\dagger} \thinspace \hat{\mathbf{R}} \rVert_2^2,
\end{equation}

where $\dagger$ denotes the Moore-Penrose pseudo-inverse of a matrix. Unlike TDSR, this approach has only one non-linear parameter to be optimized, reducing the problem to a one-dimensional search which can be robustly solved using the FMIN method by Brent \cite{Brent1973}.  This new approach of combining the VARPRO method with baseline fitting to align spectra will be referred to as RATS - Robust Alignment to a Target Spectrum.

\subsection{Correction method performance evaluation}
Simulated and acquired MRS data were both used to compare the performance of RATS and TDSR over a range of conditions. All simulations were generated from a linear combination of metabolite, lipid and macromolecule signals in proportions to match the appearance of normal appearing brain. Metabolite signals were simulated for a PRESS sequence (TE=30ms at 3T) using density matrix operators \cite{Levitt2001} and published chemical shift and J-coupling values \cite{Govind2015}. 5Hz linebroadening was applied to all simulated spectra prior to the addition of Gaussian distributed complex noise. The noise standard deviation was adjusted to produce a desired spectral SNR - defined here as the maximum metabolite spectral intensity divided by the standard deviation of the spectral noise.

\subsubsection{Simulations}
The first simulation test evaluated the frequency correction accuracy as a function of SNR and frequency shift magnitude. 512 spectra were generated, each with the same spectral signals and SNR ratio but differing random noise samples. A linearly increasing frequency shift was applied to each spectrum, where the first and last spectra had shifts of 0Hz and 10Hz respectively. Seven sets of 512 spectra were generated, with each set having one of the following SNR values: 2.5, 5.0, 7.5, 10.0, 15.0, 20.0 and 25.0. The second simulation test was identical to the first, with the exception that the phase of each spectrum is randomly altered to be between -180 and 180 degrees with a uniform probability distribution. The third simulation test evaluated the performance of each method to baseline instability originating from the tails of a residual water resonance. Similar to the second simulation test, the metabolite, lipid and macromolecule signals were increasingly shifted to a maximum value of 10Hz, over 512 randomly phased spectra with a SNR of 15. A randomly phased artificial residual water resonance, at 4.65 PPM with a linewidth of 10 Hz, was added to introduce a moderate unstable baseline artifact combined with phase and frequency perturbations.

For all simulations, frequency or phase adjustments were not applied to the first spectrum of each simulated set since it was used as the reference spectrum. The frequency and phase errors for each approach were measured by subtracting the estimated values from the true values and calculating the standard deviation for each simulation set.

\subsubsection{Edited MRS}
A glutathione (GSH) J-difference edited example dataset was used to compare the correction methods, due to its high sensitivity to spectral misalignment. GSH edited in-vivo MRS was acquired from a single healthy volunteer using a MEGA-PRESS sequence on a 3T Philips Achieva scanner (Philips Healthcare, Eindhoven, Netherlands). A 3x3x2cm voxel was placed in the anterior cingulate cortex (ACC) and 480 averages acquired with the following acquisition parameters: TR=2s; TE=131; 55Hz bandwidth editing pulse at 4.56 PPM; 1024 complex data points acquired at a sampling frequency of 2000Hz.

The correction procedure started with the calculation of the median spectrum separately for the edited and non-edited scans. The individual average closest to the median (calculated using a least-squares spectral difference) was automatically selected as the correction target, a strategy designed to reduce the chances of a motion corrupted average being used as a reference. The correction of individual averages was performed separately for edited and non-edited scans before calculation of the corrected mean edited and non-edited scan. A second correction step was performed to minimize subtraction artifacts by correcting the mean non-edited scan to match the edited (reference) scan over the tNAA spectral region (1.8 to 2.2 PPM). The same correction method (RATS or TDSR) was used for the initial and subtraction correction steps for comparison. Finally, 3Hz linebroading and zero-filling to 4096 points was applied to aid a visual comparison between the RATS, TDSR and uncorrected processing variants.

\subsubsection{Implementation details}
In the original description of the TDSR method \cite{Near2015}, a preprocessing step to restrict the spectral region for registration (using the discrete Fourier transform) was optionally performed to reduce the influence of unstable signals, such as residual water. In addition, the latter points of the time-domain signals were removed prior to optimization to reduce the noise contribution. In this comparison, the same ``preprocessing'' steps were taken, with only the spectral region between 4 and 0.5 PPM and the first 200ms of the free induction decay (FID) being considered for both methods - unless stated otherwise. For the RATS method, zero, first and second degree polynomial functions were used to construct the baseline modelling basis and a maximum frequency shift limit of $\pm 20$Hz was used for the Brent optimization algorithm.

The RATS and TDSR methods were implemented in the R programming language (v3.5.0) \cite{R2018} and integrated into the spant (SPectroscopy ANalysis Tools) package (v0.11.0) for MRS analysis (https://github.com/martin3141/spant). All code and data used to generate the results from this paper is freely available from: https://github.com/martin3141/rats.

\section{Results}
\begin{figure}
\begin{center}
\includegraphics[width=0.9\textwidth]{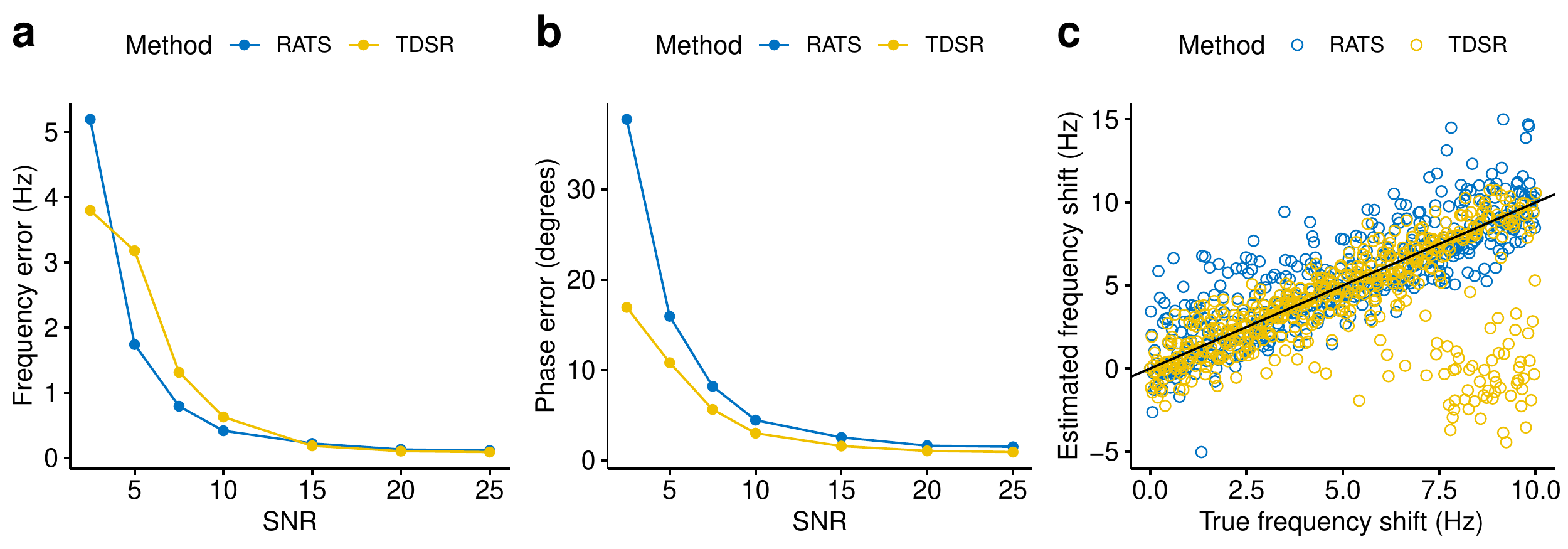}
\caption{Frequency a) and phase b) correction accuracy of RATS and TDSR for simulated spectra over a range of SNR values. Frequency was linearly increased up to 10Hz for each SNR set of 512 spectra. c) estimated vs true frequency shifts for SNR=5.}
\label{freq}
\end{center}
\end{figure}

Figure \ref{freq} compares the accuracy of RATS and TDSR for frequency shifts up to 10Hz. At the lowest SNR value of 2.5, the TDSR method shows improved frequency correction accuracy (part a) over RATS, however for the SNR range between 5 and 15 RATS is the more accurate method. Scatter plot c) illustrates how the TDSR method becomes increasing unstable for frequency shifts greater than 5Hz for SNR=5, and this is the cause of the reduced performance compared to RATS in this SNR regime. Part b) shows TDSR has improved phase correction accuracy over RATS, with both methods becoming comparable in both frequency and phase correction for SNR=15 and above.

% Figure 1

\begin{figure}
\begin{center}
\includegraphics[width=0.6\textwidth]{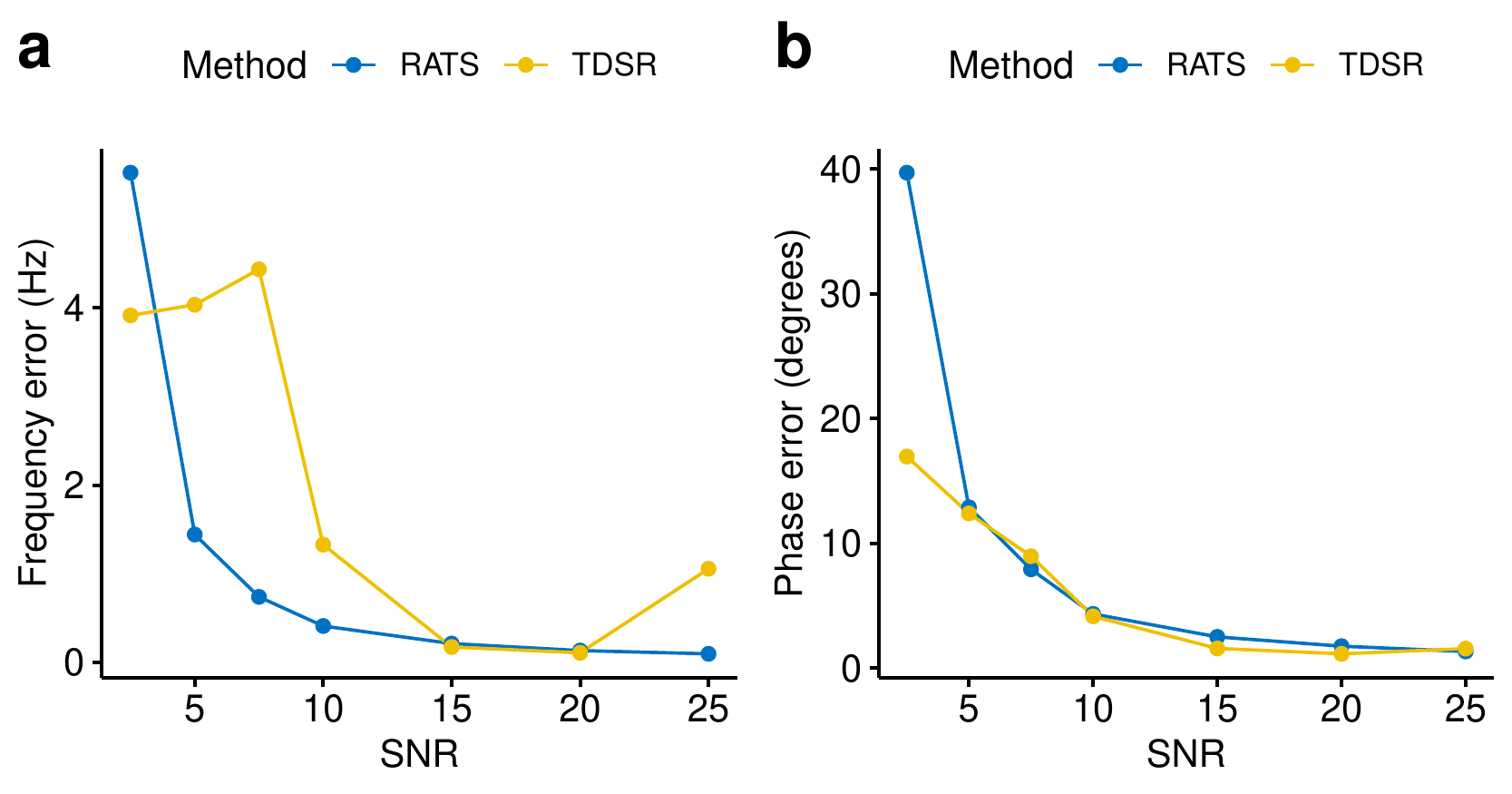}
\caption{Frequency a) and phase b) correction accuracy of RATS and TDSR for simulated spectra over a range of SNR values. Spectra were randomly phased and frequency was linearly increased up to 10Hz for each SNR set of 512 spectra.}
\label{phase}
\end{center}
\end{figure}

Figure \ref{phase} shows how RATS and TDSR perform with combined frequency and phase perturbations. Part a) shows a similar trend to Figure \ref{freq} with the RATS method proving more accurate in the SNR range between 5 and 15, however some further instability was seen for the TDSR method at SNR=25. Part b) shows how the phase correction accuracy between RATS and TDSR was much closer than the more trivial test in Figure \ref{freq}b. To summarize Figures \ref{freq} and \ref{phase}, the RATS method produces more accurate frequency correction in the SNR range between 5 and 15, whereas TDSR performs better for low SNR values of less than 5. Above a SNR of 15 both methods are comparable, however, even at high SNR, some instability was found for the TDSR method when large frequency shifts (>5Hz) were combined with phase variation.

% Figure 2

\begin{figure}
\begin{center}
\includegraphics[width=0.7\textwidth]{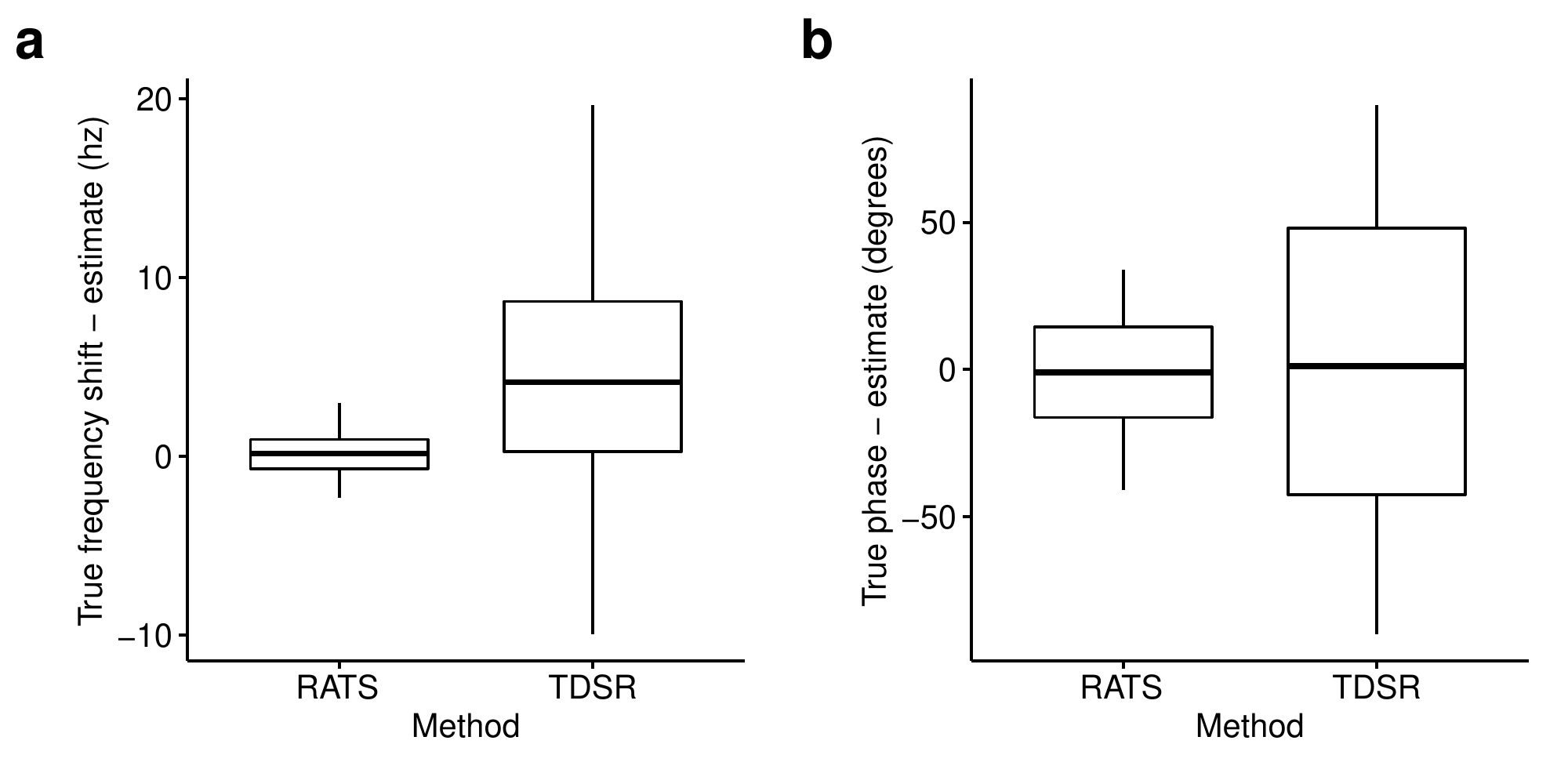}
\caption{Frequency a) and phase b) correction accuracy of RATS and TDSR for simulated spectra (SNR=15) with unstable baselines. Residual water signals were generated to simulate baseline distortion and spectra were also randomly phased with linearly increasing frequency shifts up to 10Hz over 512 spectra.}
\label{bl_perf}
\end{center}
\end{figure}

\begin{figure}
\begin{center}
% below needed for online submission
% \includegraphics[width=0.9\textwidth,natwidth=1200,natheight=500]{figure4.png}
\includegraphics[width=0.9\textwidth]{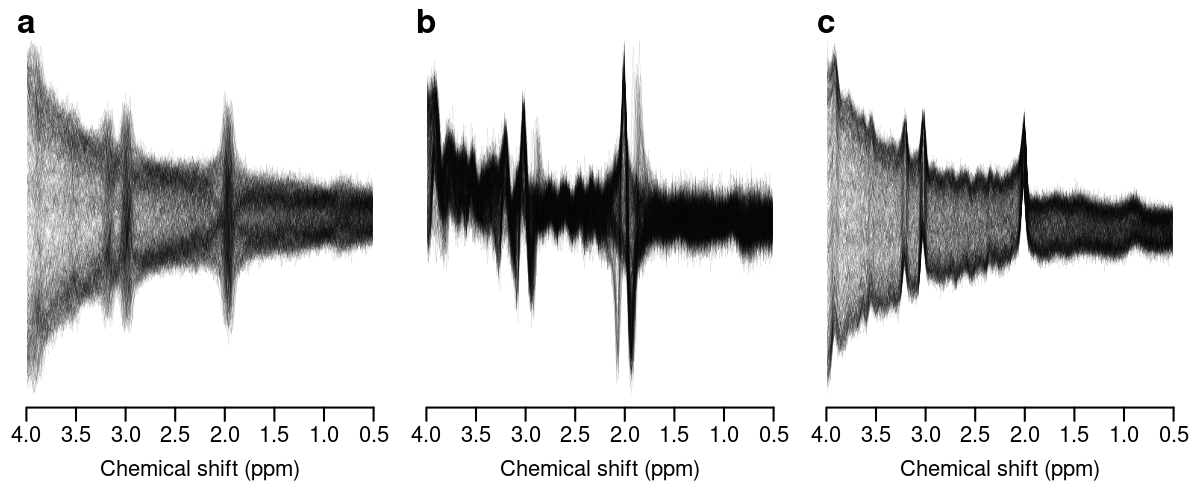}
\caption{a) 512 overlaid simulated spectra (SNR=15) with unstable baseline, random phase and linearly increasing frequency shifts up to 10Hz. b) TDSR corrected spectra and c) RATS corrected spectra.}
\label{bl_spectra}
\end{center}
\end{figure}

The performance of each method in the presence of simulated baseline instability, combined with frequency and phase perturbations, is shown in Figure \ref{bl_perf}. For stable baseline simulations (Figures \ref{freq} and \ref{phase}), a SNR of 15 was shown to produce good results for both methods, however in the unstable case the RATS method has reduced bias and improved accuracy for both frequency (Figure \ref{bl_perf} a) and phase (Figure \ref{bl_perf}b) correction. This improvement in accuracy is illustrated in Figure \ref{bl_spectra}, where RATS (part c) shows closely aligned and phased spectra compared to TDSR (part b).

% Figure 3

% Figure 4

\begin{figure}
\begin{center}
\includegraphics[width=0.9\textwidth]{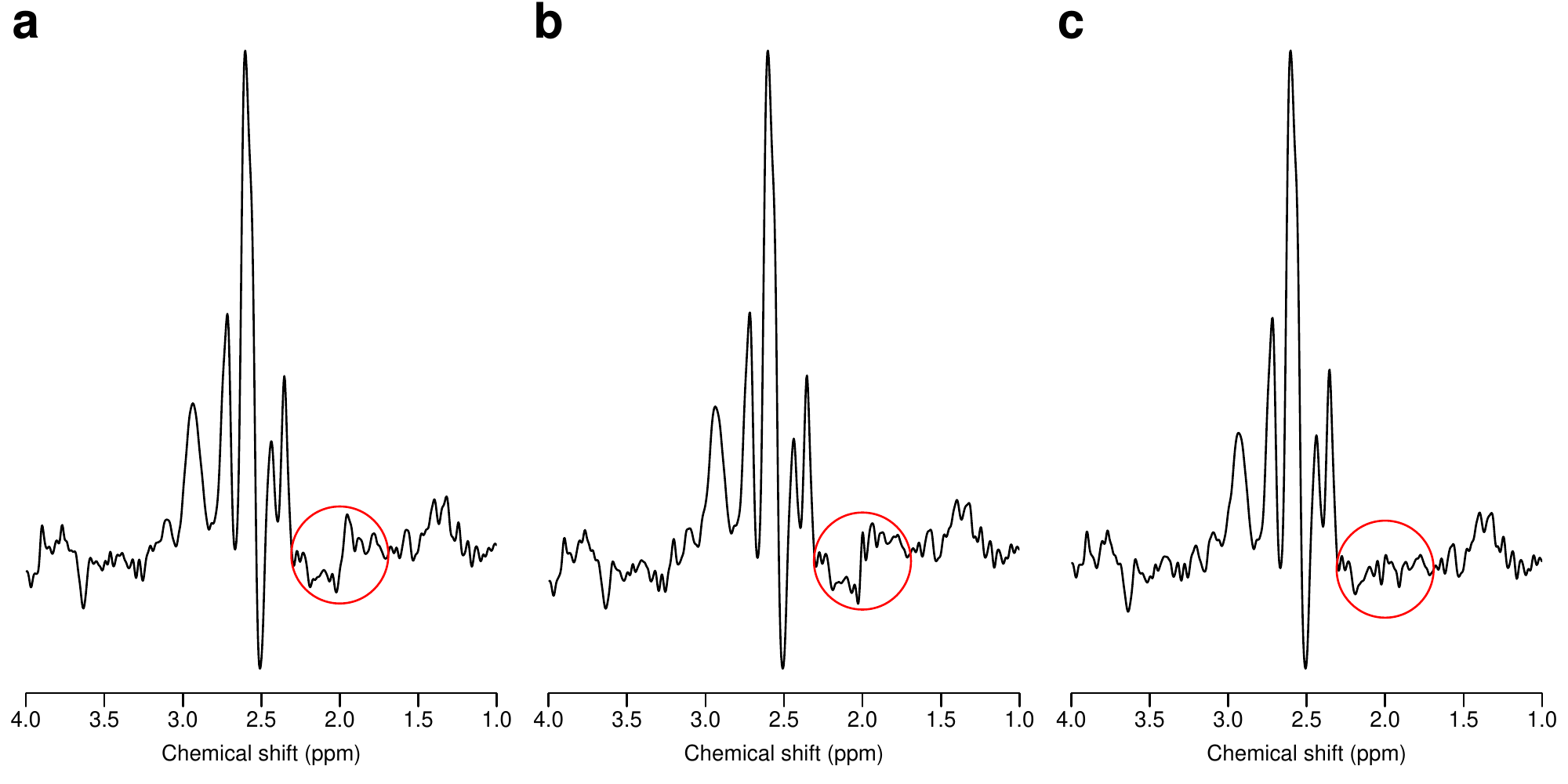}
\caption{In-vivo edited GSH spectra from a voxel placed in the ACC of a healthy participant. a) uncorrected data, b) TDSR corrected data and c) RATS corrected data. The tNAA subtraction artifact region is highlighted with a red circle.}
\label{ed_gsh_spec}
\end{center}
\end{figure}

Edited GSH spectra are shown in Figure \ref{ed_gsh_spec}, with a comparison between a) uncorrected, b) TDSR corrected and c) RATS corrected data. Frequency or phase errors between the edited and non-edited averages result in imperfect subtraction, resulting in residual signal, most clearly seen in the tNAA spectral region between 1.8 and 2.2 ppm. Moderate distortions in the tNAA are present in the uncorrected and TDSR corrected data, whereas RATS correction either eliminates these distortions or reduces them to be indistinguishable from noise. The impact of these artifacts on the edited GSH resonance at 2.95 ppm can also be seen, with uncorrected and TDSR corrected data showing erroneously elevated GSH due to an incomplete subtraction of the tCr peak.

% Figure 5

\section{Discussion and Conclusion}
In this work, the incorporation of baseline modeling and VARPRO into retrospective spectral correction have been shown to offer: 1) improved robustness to frequency shifts greater than 5Hz; 2) improved robustness to unstable baseline distortions and 3) reduced subtraction artifacts for GSH J-difference edited MRS. The improved robustness to larger frequency shifts results from to use of a VARPRO formulation to reduce the optimization complexity from two to one dimension. This allows the use of optimization methods based on a 1D search, which are less prone to converging on local-minima. In the low SNR regime (less than 5) the new method was generally less accurate that TDSR, likely resulting from the increased modeling freedom due to the addition of a baseline basis set. However, in this regime both methods performed poorly, with frequency correction errors greater than 3Hz, and therefore the use of either method may not be advisable for low SNR spectra. 

Whilst correction accuracy was the main focus of this work, it should also be noted that both RATS and TDSR correction may be performed quickly on modern hardware. For instance, the correction of 128 averages takes approximately 0.6 and 0.4 seconds for TDSR and RATS respectively using and Intel(R) Core(TM) i5-8250U CPU. Therefore, in cases where SNR is low and the best method may not be obvious, it is feasible to compare the SNR from averaging uncorrected; TDSR and RATS corrected data and proceed with the highest quality reconstruction.

One alternative to spectral-registration based methods is known as the metabolite-cycling technique \cite{Dreher2005}, where metabolite selective inversion pulses are alternately applied prior to the localization scheme as an alternative to conventional water suppression. Using this scheme, a full intensity water signal is acquired for each average and water-suppressed metabolite data may be obtained by subtracting average pairs. Therefore, accurate phase and frequency correction may be performed using the high SNR water signal in protocols where the metabolite SNR may be too low for spectral registration with TDSR or RATS \cite{Hock2013, Doering2018}. Whilst effective for low metabolite SNR applications, at the time of writing the metabolite-cycling method is not widely available or suitable for non-proton MRS.

The first reported use of retrospective correction for conventional clinical MRS was in 2005 \cite{Oz2005}. Yet, despite being compatible with all widely available sequences and rapid to perform, current use remains largely restricted to edited-MRS \cite{Evans2013}. One potential reason may be due the smaller typical voxel dimensions used for conventional clinical MRS (2cm sided cube) compared to edited MRS (3cm sided cube) resulting in a metabolite SNR lower than required for accurate correction. However, previous work \cite{Oz2005} has shown that combining averages over blocks is effective for using spectral registration with lower SNR data. Further potential barriers to use in the clinical setting include the extra time required to export individual averages for offline analysis and limited availability of spectral registration methods integrated into the scanner software. Recently available open-source implementations of spectral registration methods in the MATLAB (MathWorks, Natick, Massachusetts, USA) based FID-A package \cite{Simpson2017} or R \cite{R2018} based spant package (https://github.com/martin3141/spant) may aid clinical validation and uptake in the future.

Whilst the focus of this paper is on the correction of distorted scans, the RATS method produces an amplitude, frequency offset and phase offset for each average, which may be also used as criteria for excluding individual averages from the final result. For instance, a frequency offset greater than 5Hz could act as an exclusion criterion for a particular average, and a dataset with more than 10\% of averages excluded may indicate significant movement which should be incorporated into the clinical decision making process. Plots of the amplitude, frequency and phase throughout the scan could also accompany the fitting results to aid quality assessment and clinical interpretation.

In conclusion, the RATS algorithm has been shown to provide accurate retrospective correction of SVS MRS data in the presence of large frequency shifts and baseline instability. The method may be easily incorporated into the processing pipeline of both conventional and J-difference edited MRS to improve lineshape, SNR and aid quality assessment.

\section*{ACKNOWLEDGMENTS}
The support of Philips Healthcare Clinical Science for the provision of the MEGA-PRESS implementation.

%\section*{CONFLICT OF INTEREST}

% Submissions are not required to reflect the precise reference formatting of the journal (use of italics, bold etc.), however it is important that all key elements of each reference are included.
\bibliography{main}

%\graphicalabstract{example-image-1x1}{Please check the journal's author guildines for whether a graphical abstract, key points, new findings, or other items are required for display in the Table of Contents.}

\end{document}